\documentclass[letterpaper, 10 pt, conference]{ieeeconf}  

\IEEEoverridecommandlockouts                              

\usepackage[hyphens]{url}
\usepackage{amsmath}
\usepackage{amsfonts}
\usepackage{amssymb}
\usepackage{bm}
\usepackage{cite}
\usepackage{color}
\usepackage[pdftex]{graphicx}
\renewcommand{\baselinestretch}{0.9745}

\title{\LARGE \bf
Sharing the Load: Considering Fairness in De-energization Scheduling to Mitigate Wildfire Ignition Risk using Rolling Optimization\vspace*{-0.4em}%
}

\author{Alyssa Kody,$^{1}$ Amanda West,$^{2}$ and Daniel K. Molzahn$^{2}\vspace*{-0.4em}$
\thanks{$^{1}$Energy Systems Division, Argonne National Laboratory, {\tt\small akody@anl.gov}. 
This work is partially supported by the U.S. Department of Energy, Office of Electricity and Laboratory Directed Research and Development (LDRD) funding from Argonne National Laboratory, provided by the Director, Office of Science, of the U.S. Department of Energy under Contract No. DE-AC02-06CH11357.
}%
\thanks{$^{2}$School of Electrical and Computer Engineering, Georgia Institute of Technology, {\tt\small \{awest93,molzahn\}@gatech.edu}. 
This work is partially supported by the U.S. National Science Foundation Energy, Power, Control and Networks (EPCN) under award \#2145564.
}%
}

\begin{document}

\maketitle
\thispagestyle{empty}
\pagestyle{empty}

\begin{abstract}
Wildfires are a threat to public safety and have increased in both frequency and severity due to climate change. To mitigate wildfire ignition risks, electric power system operators proactively de-energize high-risk power lines during ``public safety power shut-off'' (PSPS) events. Line de-energizations can cause communities to lose power, which may result in negative economic, health, and safety impacts. Furthermore, the same communities may repeatedly experience power shutoffs over the course of a wildfire season, which compounds these negative impacts. However, there may be many combinations of power lines whose de-energization will result in about the same reduction of system-wide wildfire risk, but the associated power outages affect different communities. Therefore, one may raise concerns regarding the fairness of de-energization decisions. Accordingly, this paper proposes a 
framework to select lines to de-energize in order to balance wildfire risk reduction, total load shedding, and fairness considerations. The goal of the framework is to prevent a small fraction of communities from disproportionally being impacted by PSPS events, and to instead more equally share the burden of power outages. For a geolocated test case in the southwestern United States, we use actual California demand data as well as real wildfire risk forecasts to simulate PSPS events during the 2021 wildfire season and compare the performance of various methods for promoting fairness. Our results demonstrate that the proposed formulation can provide significantly more fair outcomes with limited impacts on system-wide performance.
\end{abstract}

\section{Introduction}\label{sec:intro}




Electric power systems have ignited a number of severe wildfires over the last five years~\cite{keeley2019twenty,psps_cpuc}. To reduce to the risk of wildfire ignition, power system operators execute  ``Public Safety Power Shutoff'' (PSPS) events that temporarily de-energize power lines that run through areas with high ignition risk.
Since de-energized lines cannot ignite wildfires, PSPS strategies are effective at immediately reducing the ignition risks posed by an electric power system.
However, the partially de-energized system may be incapable of supplying all load demands, leading to power outages that have negative economic, health, and safety impacts~\cite{WSJ_cost, NPR_medical, CalMatters_lost_food}.

There may be many possible combinations of de-energized power lines that result in about the same reduction of system-wide wildfire risk, but the associated power loss affects different communities, thus raising fairness concerns with PSPS events. 
If the same communities repeatedly experience power shutoffs over the course of a wildfire season, repercussions may compound, especially for vulnerable groups \cite{abatzoglou2020population}. To address these fairness concerns, this paper proposes a rolling optimization framework to compute PSPS de-energization decisions that result in more fair power outages.

Typically, utilities de-energize all lines that exceed a pre-determined risk threshold, which is based on local ambient environmental conditions (e.g., foliage flammability, wind speed, rainfall, and temperature) and infrastructure characteristics (e.g., voltage level and line conditions, lengths, and geometries)~\cite{waseem2021,vazquez2022, rhodes2020balancing}. These are localized approaches to determining PSPS events. To achieve operators' goals of making PSPS events ``smaller in scope, shorter in duration, and smarter in performance''~\cite{PGE_WildfirePlan}, current research is considering PSPS events on the \emph{system level} with local data input. Notable recent research by Rhodes, Ntaimo, and Roald~\cite{rhodes2020balancing} proposed \emph{optimized} PSPS events.
They formulate a mixed-integer linear program (MILP) to select the set of power lines to de-energize that jointly minimizes the total wildfire ignition risk posed by the network and the system-wide load shedding.
Results from~\cite{rhodes2020balancing} suggest that the optimized PSPS formulation allows operators to tradeoff wildfire risk and system-wide load shedding, and potentially improve performance relative to a threshold-based approach. 
Building on this prior work, subsequent research has considered alternate wildfire risk models~\cite{waseem2021} and infrastructure investment plans to support optimized PSPS events~\cite{irep2022}. 

Optimized PSPS formulations provide the groundwork for explicitly considering fairness when making line de-energization decisions. Fairness is an increasingly studied topic in many fields, such as artificial intelligence~\cite{richardson2021} and wireless spectrum allocation~\cite{peng2006}, as well as in other power systems applications, e.g., energy storage~\cite{tarekegne2021}, distribution network reconfiguration~\cite{hettle2021}, transmission loss allocation~\cite{exposito2000}, and tariff design~\cite{kahn2021}. 
However, incorporating notions of fairness into PSPS decision-making introduces modeling, algorithmic, and computational challenges.

From a modeling perspective, there are many possible fairness metrics~\cite{gupta2020} and the temporal scale is also important (e.g., should one consider power outages over an hour, the course of a day, a wildfire season, or over many seasons?). 
In this exploratory work, we focus on fairness in load shedding per bus. While our proposed framework is more general, our numerical studies focus on the most severe period of a wildfire season.
We should also avoid perverse incentives that, e.g., promote unnecessary load shedding. To address these modeling challenges, we focus on a selected subset of methods for promoting more fair de-energization decisions.

From an algorithmic perspective, many plausible methods for promoting fairness depend on both historical and future de-energization decisions. These features introduce multiple time dependencies and the need to address uncertainties from fluctuations in future wildfire risks and load demands. 
We propose a rolling optimization framework that produces daily decisions for line de-energizations informed by the past load shedding at each location along with forecasts for the next day of wildfire risks and load demands. 
This framework chooses line de-energizations to simultaneously improve upon a fairness metric while minimizing wildfire ignition risks and system-wide load shedding.
After implementing the de-energization decisions 
and observing the results, the framework rolls forward to determine the decisions for the next day. This framework thus addresses both time dependencies and forecast uncertainties.

From a computational perspective, the MILP formulation used to determine line de-energization decisions can be computationally challenging.
This wildfire switching formulation is related to existing \emph{optimal transmission switching} problems that permit de-energizing transmission lines to reduce generation costs by managing network congestion during normal operations~\cite{fisher2008optimal,hedman2011survey}. The computational difficulty of optimal transmission switching problems has motivated the development of many heuristic solution techniques~\cite{fuller2012,johnson2021}. However, these heuristics are not well suited for our problem since the wildfire formulation de-energizes lines solely to reduce ignition risk rather than managing network congestion to achieve some other objective. In this exploratory work, we thus focus on short (one-day) time horizons and moderate-size networks to demonstrate the capabilities of the proposed framework. Our ongoing work aims to develop transmission switching heuristics for the wildfire setting to scale the proposed framework to large systems and longer time horizons.


To summarize, 
we propose and analyze a rolling optimization framework that solves MILP problems daily to determine line de-energization decisions. The framework provides the flexibility to consider both system-wide concerns like the wildfire ignition risk and the total load shedding as well as fairness considerations. We demonstrate this framework using a standard test system (RTS-GMLC~\cite{barrows2019ieee}) with actual wildfire risk and load demand data for the 2021 wildfire season. The results show that the proposed rolling optimization framework can provide significantly more fair outcomes with limited impacts on system-wide performance. We also observe that different methods for promoting fairness can greatly affect both the line de-energization decisions and the load shedding at each bus. Finally, the results indicate that some methods for promoting fairness may introduce perverse incentives that can lead to unnecessary load shedding.

The remainder of this paper is organized as follows. Section~\ref{sec:psps_formulation} formulates the optimized PSPS problem and our proposed rolling optimization framework. Section~\ref{sec:fairness} modifies the rolling optimization framework to incorporate fairness considerations. Section~\ref{sec:test_case} describes our test case setup. Section~\ref{sec:results} provides our numerical results. Section~\ref{sec:conclusion} concludes the paper and discusses future work.

\section{Public Safety Power Shutoffs via a Rolling Optimization Framework}
\label{sec:psps_formulation}

After introducing notation, this section describes the optimized PSPS problem formulation and our proposed rolling optimization framework that repeatedly solves this problem.

\subsection{Parameter and variable definitions}

We first present the parameters and variables associated with network operations during PSPS events.
Let $\mathcal{N}$, $\mathcal{L}$, and $\mathcal{G}$ be the sets of buses, lines, and generators, respectively. 
Let the relevant period of the wildfire season contain $J$ total days, and let index $j$ indicate the considered day in the wildfire period.
Let $\mathcal{T} = \{1, \ldots, T \}$ be the considered set of time intervals during each day, where $T$ is the final time. 
We use a per unit (p.u.) base power of 100~MVA. 

We model the network with the $B\Theta$ representation of the DC power flow approximation. This approximation neglects reactive power dispatch, line losses, and voltage magnitude variation.
Specified parameters for each line $\ell \in \mathcal{L}$ are:
\begin{itemize}
    \item $b^\ell$, line susceptance in p.u.,
    \item $\overline{f}^\ell$, the power flow limit in p.u.,
    \item $r^\ell$, the wildfire risk incurred if line $\ell$ is energized,
    \item $n^{\ell, \text{to}}$ and $n^{\ell, \text{fr}}$, \textit{to} and \textit{from} buses, respectively, where positive power flows from the \textit{from} bus to the \textit{to} bus,
    \item $\overline{\delta}^\ell$ and $\underline{\delta}^\ell$, upper and lower voltage angle difference limits, respectively, in radians.
\end{itemize}
For all generators $i \in \mathcal{G}$:
\begin{itemize}
    \item $\overline{g}^i$ and $\underline{g}^i$, upper and lower power generation limits, respectively, in p.u.,
    \item $n^i$, the bus at which generator $i$ is located.
\end{itemize}
For all buses $n \in \mathcal{N}$:
\begin{itemize}
    \item $d^n_{t} \geqslant 0$, predicted power demand at time interval $t \in \mathcal{T}$ in p.u.,
    \item $\mathcal{G}^n$, the set of generators located at bus $n$,
    \item $\mathcal{L}^{n, {\text{to}}}$ and $\mathcal{L}^{n, {\text{fr}}}$, the subset of lines $\ell \in \mathcal{L}$ with bus $n$ as the designated \textit{to} bus, and bus $n$ as the designated \textit{from} bus, respectively.
\end{itemize}
The operation of the network during a PSPS event is characterized by the following set of variables:
\begin{itemize}
    \item $g^i_{t}$, power generated in p.u. at unit $i \in \mathcal{G}$ during time interval $t \in \mathcal{T}$,
    \item $\theta^n_{t}$, voltage angle in radians at bus $n \in \mathcal{N}$ during time interval $t \in \mathcal{T}$,
    \item $s^n_{t}$, load shedding in p.u. for bus $n \in N$ during time interval $t \in \mathcal{T}$,
    \item $f^\ell_{t}$, power flowing in p.u. from bus $n^{\ell, \text{fr}}$ to bus $n^{\ell, \text{to}}$ along line $\ell \in \mathcal{L}$ during time interval $t \in \mathcal{T}$,
    \item $z^\ell \in \{0,1\}$, energization state of line $\ell \in \mathcal{L}$, where $z^\ell=0$ indicates line $\ell$ is de-energized and  $z^\ell=1$ indicates line $\ell$ is energized.
\end{itemize}

\subsection{Operational and physical constraints}

The outputs of all generators $i \in \mathcal{G}$ must be within their lower and upper limits during each time period:
\begin{align} \label{const: gen limits}
    \underline{g}^i \leqslant g^i_{t} \leqslant \overline{g}^i, \qquad \forall i \in \mathcal{G}, \ \forall t \in \mathcal{T}.
\end{align}
The load shed at all buses and for all time instances must be positive and cannot exceed the power demand at the bus:
\begin{align} \label{const: load shed limits}
    0 \leqslant s^n_{t} \leqslant d^n_{t}, \quad \forall n \in \mathcal{N}, \ \forall t \in \mathcal{T}.
\end{align}
The power flow along line $\ell \in \mathcal{L}$ must not exceed upper and lower line flow limits; however, if de-energized, then the power flow along line $\ell$ is zero:
\begin{align} \label{const: powr flow limits}
    -\overline{f}^\ell z^\ell \leqslant f^\ell_{t} \leqslant \overline{f}^\ell z^\ell, \quad \forall \ell \in \mathcal{L}, \ \forall t \in \mathcal{T}.
\end{align}
We require that the voltage angle differences for all energized lines satisfy lower and upper limits:
\begin{multline} \label{const: voltage angle}
    \overline{\delta}^\ell z^\ell + \overline{M}(1-z^\ell) \leqslant \theta^{n^{\ell, \text{fr}}}_{t} - \theta^{n^{\ell, \text{to}}}_{t} \leqslant \underline{\delta}^\ell z^\ell + \underline{M}(1-z^\ell)\\
    \qquad \forall \ell \in \mathcal{L}, \ \forall t \in \mathcal{T},
\end{multline}
where $\overline{M}$ and $\underline{M}$ are big-M constants. We compute these constants by summing the angle difference bounds across all lines, but note that more sophisticated approaches (e.g.,~\cite{li2021}) could lead to faster solution times.
The powers flows on all lines during all time periods are governed by:
\begin{multline} \label{const: power flow}
    -b^\ell(\theta^{n^{\ell, \text{fr}}} - \theta^{n^{\ell, \text{to}}}) + |b^\ell|\underline{M}(1-z^\ell) \leqslant f_t^\ell\\
    \leqslant -b^\ell(\theta^{n^{\ell, \text{fr}}} - \theta^{n^{\ell, \text{to}}}) + |b^\ell|\overline{M}(1-z^\ell),\\
    \forall \ell \in \mathcal{L}, \ \forall t \in \mathcal{T}.
\end{multline}
Last, we require power balance at all buses:
\begin{multline} \label{const: power balance}
    \sum_{\ell \in \mathcal{L}^{n, \text{fr}}} f^\ell_{t} - \sum_{\ell \in \mathcal{L}^{n, \text{to}}} f^\ell_{t} = s^n_{t} - d^n_{t} + \sum_{i \in \mathcal{G}^n} g^i_{t}, \\
    \forall n \in \mathcal{N}, \ \forall t \in \mathcal{T}.
\end{multline}

\subsection{Load shedding and wildfire risk reduction objective}

Our objective is to minimize both load shedding and wildfire risk reduction. These objectives often conflict since de-energizing lines reduces wildfire risk but typically increases the difficulty of meeting the load demands.
Let $\alpha \in [0,1]$ be a parameter that quantifies the operator's priority between minimizing load shed and minimizing wildfire risk, with large values of $\alpha$ prioritizing load shedding and small values of $\alpha$ prioritizing wildfire ignition risk.

Let $D$ be the total predicted demand in the network, i.e.,
$D = \sum_{t \in \mathcal{T}}\sum_{n \in \mathcal{N}} d^n_{t}$,
and let $R$ be the total predicted wildfire risk of the network if all lines $\ell \in \mathcal{L}$ are energized, i.e.,
$R = \sum_{\ell \in \mathcal{L}} r^\ell$.
Let $C(\cdot)$ be the objective function that the user wishes to minimize, which is a function of all load shedding values ($s$) and de-energization decisions ($z$):
\begin{multline} \label{objective func}
    C(s, z) = \frac{\alpha}{D}\left(\sum_{t \in \mathcal{T}}\sum_{n \in \mathcal{N}} s^n_{t}\right)  + \frac{(1-\alpha)}{R}\left(\sum_{\ell \in \mathcal{L}} r^\ell z^\ell\right).
\end{multline}

\subsection{Rolling optimization framework}

Given parameters $\alpha$ and $T$, we formulate the PSPS problem for wildfire day $j$ as:
\begin{equation}
\begin{aligned}
&\min\limits_{g, \theta, f, s, z} \ \eqref{objective func} \ \ \text{s.t.} \ \eqref{const: gen limits} - \eqref{const: power balance}. 
\end{aligned}
\label{PSPS}
\end{equation}

We simulate the specified period of de-energization decisions using rolling optimization by executing the following:
\begin{enumerate}
    \item Initialize $j=1$ as the first day in the wildfire period.
    \item For the $j^{th}$ day of the wildfire period, retrieve:
    \begin{itemize}
        \item Wildfire risk predictions $r^{\ell}$ for all lines $\ell \in \mathcal{L}$,
        \item Demand predictions $d^n_{t}$ for all buses $n \in \mathcal{N}$ and time intervals $t \in \mathcal{T}$,
        \item Tradeoff parameter $\alpha$.
    \end{itemize}
    \item Solve optimization problem \eqref{PSPS}.
    \item During the $j^{th}$ day in the wildfire period, de-energize all lines for which $z^\ell=0$.
    \item Update day index: $j \leftarrow j+1$
    \item If $j>J$, end simulation, else repeat from step (2).
\end{enumerate}

\section{Consideration of Fairness}
\label{sec:fairness}


There are many possible notions of fairness~\cite{gupta2019individual,gupta2020}. 
To maintain computational tractability in this exploratory work, we focus on methods for which \eqref{PSPS} remains a MILP when modified to consider fairness. 
We intend to analyze more general alternatives in future work.
This section first formulates the PSPS rolling optimization framework using a generic fairness consideration method and then describes three different methods for incorporating fairness.

\subsection{Considering fairness in PSPS optimization problems}

We consider three methods for promoting fairness of load shedding within the rolling optimization framework.
Each method requires a different function to measure fairness and different optimization constraints. 
Therefore, in this subsection, we first introduce a generic fairness function that we will later explicitly define for each considered fairness method.
Let $F_j(\cdot)$ be a function valued between 0 and 1 (inclusive) that maps from 
optimization variable inputs to a measure of load shedding fairness for the $j^{th}$ day in the wildfire period.
We define $F_j(\cdot)=1$ as the \textit{least fair} outcome, and $F_j(\cdot)=0$ as the \textit{most fair} outcome.
Note that the fairness mapping $F_j(\cdot)$ changes depending on the day $j$ in the wildfire period to account for prior load shedding. 
Via mild abuses of notation, the generic fairness function $F_j(\cdot)$ 
will be replaced in the subsequent subsections.

Enforcing \textit{fair} load shedding decisions may increase wildfire ignition risk relative to solutions that do not consider fairness.
To ensure that wildfire ignition risk does not significantly increase, we restrict the total wildfire risk of the fairness-consideration version of the de-energization problem to be no greater than a small percentage above the solution obtained without considering fairness.
Let $\hat{z}^\ell$, $\forall \ell \in \mathcal{L}$ be the de-energization decisions found from solving problem \eqref{PSPS} for the $j^{th}$ day of the wildfire period.
Let $\zeta>0$ be the allowable fraction of wildfire risk over the version without fairness consideration.
We constrain the wildfire risk as:
\begin{align}\label{const: limit risk}
    & \sum_{\ell \in \mathcal{L}} r^\ell z^\ell \leqslant (1+\zeta)  \sum_{\ell \in \mathcal{L}} r^\ell \hat{z}^\ell.
\end{align}

Since we have restricted the total amount of risk the network can pose, our objective is now to simultaneously minimize overall load shedding and the fairness function $F_j(\cdot)$.
However, these are often competing objectives.
For example, a completely ``fair'' outcome could be that all buses have none of their demand met, i.e., all load in the network is shed.
Obviously, this is not a desirable result.
Another completely ``fair'' outcome is that all buses have all their demands met fully, but given that we require a certain reduction in wildfire risk via line de-energization, it is likely that load shedding will sometimes be necessary.

Let $\beta \in [0,1]$ be a parameter that quantifies the operator's priority between minimizing total load shed and minimizing the fairness function, with large values of $\beta$ prioritizing total load shed and small values of $\beta$ prioritizing fairness.
Let $C_j^{\text{fair}}(\cdot)$ be the objective function that mathematically formulates this tradeoff for the $j^{th}$ day in the wildfire period:
\begin{multline} \label{fair objective func}
    C_j^{\text{fair}}(s, S_{\text{max}}, S_{\text{min}}) = \frac{\beta}{D}\left(\sum_{t \in \mathcal{T}}\sum_{n \in\mathcal{N}} s^n_{t}\right)\\
    + (1-\beta)F_j(s, S_{\text{max}}, S_{\text{min}}),
\end{multline}
where the fairness function $F_j(\cdot)$ is analogous to a regularizing term 
and, as defined below, $S_{\text{max}}$ and $S_{\text{min}}$ are optimization variables that designate the maximum and minimum discounted
load shed over the wildfire period experienced at any
bus $n \in \mathcal{N}$.
In \eqref{fair objective func}, we let the general fairness mapping $F_j(\cdot)$ be a function of $s$, $S_{\text{max}}$ and $S_{\text{min}}$; however, some of these inputs will be extraneous when we replace $F_j(\cdot)$ with specific notions of fairness in Sections~\ref{sec:min max load shed}, \ref{weight}, and \ref{range}.

Let $\check{s}^n_{(m,t)}$ be the actual load shed during real-time operation at bus $n \in \mathcal{N}$ and time $t\in \mathcal{T}$ for day $m$ of the wildfire period. Define $S^n_{j}$ to be the running tally of the actual load shed at bus $n \in \mathcal{N}$ from day $1$ to $j-1$ of the wildfire period:
\begin{align}\label{past_load_shed}
    S^n_{j} = \begin{cases} 0, & j=1\\
    \sum\limits_{m=1}^{j-1} \eta^{(j-1-m)} \sum\limits_{t \in \mathcal{T}} \check{s}^n_{(m,t)}, & j>1,
    \end{cases}
\end{align}
where $\eta \in [0,1]$ is a forgetting factor.
Then:
\begin{align}
    S_{\text{max}} \geqslant S^n_{j} +  \sum\limits_{t \in \mathcal{T}}  s^n_{t}, \quad \forall n \in \mathcal{N} \label{min_max_constraints}\\
    S_{\text{min}} \leqslant S^n_{j} +  \sum\limits_{t \in \mathcal{T}} s^n_{t}, \quad \forall n \in \mathcal{N}. \label{range_min} 
\end{align}%

Given parameters $\hat{z}$, $\beta$, $\zeta$, and $T$, we can formulate the PSPS problem with fairness considerations for the $j^{th}$ day of the wildfire period as:
\begin{equation}
\begin{aligned}
&\min\limits_{g, \theta, f, s, z, S_{\text{max}}, S_{\text{min}} } \ \eqref{fair objective func} \ \ \text{s.t.} \ \eqref{const: gen limits} - \eqref{const: power balance}, \eqref{const: limit risk}, \eqref{min_max_constraints}, \eqref{range_min}. 
\end{aligned}
\label{PSPS-Fair}
\end{equation}
Depending on which fairness method is implemented (see Sections \ref{sec:min max load shed} through \ref{range}), variables $S_{\text{max}}$ and $S_{\text{min}}$ and constraints \eqref{min_max_constraints} and \eqref{range_min} may not be necessary, but we have included them here to formulate a generic optimization problem applicable to all considered notions of fairness.

\begin{figure}
  \centering
  \vspace*{0.5em}
  \includegraphics[width=1\linewidth]{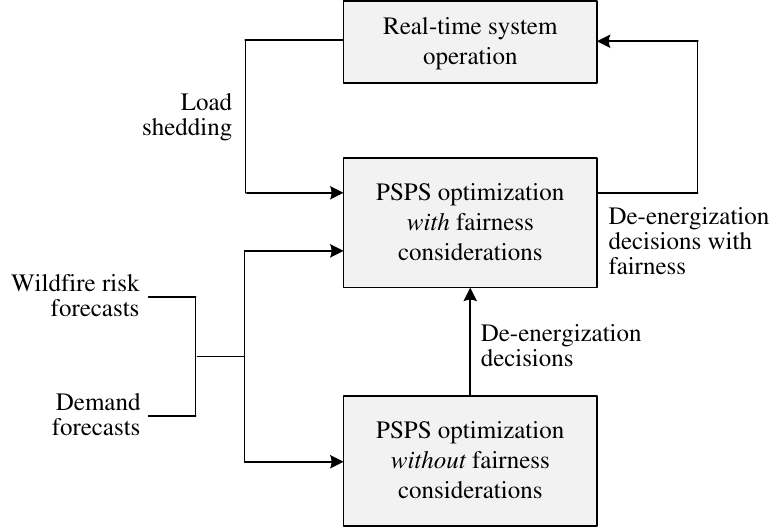}
  \caption{PSPS rolling optimization framework with fairness considerations.}
  \label{figure:feeback loop}
  \vspace*{-1em}
\end{figure}

After determining de-energization decisions, the selected lines are de-energized and the network is operated in real time.
Figure~\ref{figure:feeback loop} shows the PSPS rolling optimization scheme with fairness considerations, consisting of these steps:
\begin{enumerate}
    \item Initialize $j=1$ as the first date in the wildfire period and initialize the fairness function $F_j(\cdot)$.
    \item For the $j^{th}$ date in the wildfire period, retrieve:
    \begin{itemize}
        \item Wildfire risk predictions $r^{\ell}$ for all lines $\ell \in \mathcal{L}$,
        \item Demand predictions $d^n_{t}$ for all buses $n \in \mathcal{N}$ and time intervals $t \in \mathcal{T}$,
        \item Tradeoff parameter $\alpha$.
    \end{itemize}
    \item Solve the optimization problem \eqref{PSPS} and retrieve de-energization decisions $\hat{z}$.
    \item Solve the optimization problem with fairness consideration \eqref{PSPS-Fair}, using $\hat{z}$ to constrain the total allowable wildfire ignition risk.
    \item During the $j^{th}$ day of the wildfire period, de-energize all lines for which $z^\ell=0$. Operate the network in real time, and retrieve the actual load shedding experienced in the network $\check{s}$.
    \item Update the day index $j \leftarrow j+1$ and the fairness function $F_j(\cdot)$ based on load shedding from real-time operation.
    \item If $j>J$, end simulation, else repeat from step (2).
\end{enumerate}

\subsection{Minimize maximum load shed}\label{sec:min max load shed}

The first fairness method we consider is minimizing the maximum load shed per bus over the course of the wildfire period.
Let $F_j^{\text{max}}(\cdot)$ be defined as:
\begin{align}
    F_{j}^{\text{max}}(S_{\text{max}}) = \frac{S_{\text{max}} - \max\limits_{n \in \mathcal{N}}\{S_j^n\}}{\max\limits_{n \in \mathcal{N}} \left\{ S^n_j +  \sum\limits_{t \in \mathcal{T}} d^n_{t} \right\} - \max\limits_{n \in \mathcal{N}}\{S_j^n\}}, 
\end{align}
where the terms with maximums are used to scale the function such that feasible outputs are in the range $[0,1]$. 
Note that these terms are functions of parameters, not variables, and thus $F_{j}^{\text{max}}(\cdot)$, as defined above, is linear in the variable $S_{\text{max}}$.
The denominator is non-zero since there will always be some demand $d > 0$. 
When considering this fairness method, we solve optimization problem \eqref{PSPS-Fair} with $F_j(\cdot) \leftarrow F_j^{\text{max}}(\cdot)$ in objective function \eqref{fair objective func}. Note that the variable $S_{\text{min}}$ and the constraint \eqref{range_min} are unnecessary in this formulation.

\subsection{Weighted load shedding penalties} \label{weight}

The second fairness method penalizes load shed at bus~$n$ by the past cumulative weighted load shed at that bus, $S^n_j$, as defined in \eqref{past_load_shed}.
For this method, the relevant variable for measuring fairness is the load shed per bus.
We measure fairness via the function $F_j^{\text{weight}}(\cdot)$, which we define as:
\begin{align}
    F_j^{\text{weight}}(s) = \begin{cases} 0, & j =1\\
    \frac{\sum\limits_{n \in \mathcal{N}} \sum\limits_{t \in \mathcal{T}} S^n_{j} s^n_{t}}{\sum\limits_{n \in \mathcal{N}} \sum\limits_{t \in \mathcal{T}} S^n_{j} d^n_{t}}, & j>1,
    \end{cases} 
\end{align}
where the denominator for the $j>1$ case is a normalization term.
For $j=1$, the first day of the wildfire period, there has not been any past load shed in the network. Therefore, we define $F_1^{\text{weight}}(s) = 0$.
We solve optimization problem \eqref{PSPS-Fair} with $F_j(\cdot) \leftarrow F_j^{\text{weight}}(\cdot)$ in objective function \eqref{fair objective func}. The variables $S_{\text{max}}$ and $S_{\text{min}}$ and the constraints \eqref{min_max_constraints} and \eqref{range_min} are unnecessary in this formulation.

\subsection{Minimize load shed range}\label{range}

In the third considered fairness method, we aim to minimize the range of load shed, i.e., the difference between $S_{\text{max}}$ and $S_{\text{min}}$.
We first calculate the maximum and minimum possible values of the range $(S_{\text{max}} - S_{\text{min}})$.
The maximum possible range, $w_{\text{max}}$,~is:
\begin{align}
    w_{\text{max}} = \max\limits_{n \in \mathcal{N}} \left\{ S^n_j +  \sum\limits_{t \in \mathcal{T}} d^n_{t} \right\} - \min\limits_{n \in \mathcal{N}^d}\{S_j^n\}, \nonumber
\end{align}
where $\mathcal{N}^d \in \mathcal{N}$ denotes buses that have non-zero demand:
$\mathcal{N}^d = \{ n \in \mathcal{N} \ | \ d^n_{t} > 0, \ \forall t \in \mathcal{T} \}$.
Let $w_{\text{min}}$ be the minimum possible range:
 \begin{equation*}
    w_{\text{min}} = \max \Big\{0, \ \Big(\max_{n \in \mathcal{N}}\{S_j^n\} -
    \min\limits_{n \in \mathcal{N}^d} \left\{ S^n_j +  \sum\nolimits_{t \in \mathcal{T}} d^n_{t} \right\} \Big) \Big\}. \nonumber
\end{equation*}    
We define the associated fairness function $F_j^{\text{range}}(\cdot)$ as:
\begin{align}
    F_{j}^{\text{range}}(S_{\text{min}},  S_{\text{max}}) = \frac{(S_{\text{max}} - S_{\text{min}}) - F_{\text{min}}^{\text{range}}}{F_{\text{max}}^{\text{range}} - F_{\text{min}}^{\text{range}}}, 
\end{align}
where the denominator is nonzero assuming that there is always some demand in the network.
We solve problem \eqref{PSPS-Fair} with $F_j(\cdot) \leftarrow F_j^{\text{range}}(\cdot)$ in the objective function~\eqref{fair objective func}.
\section{Test Case Description}
\label{sec:test_case}

To demonstrate our proposed framework, we consider the ten-day period from June 13, 2021 to June 22, 2021. This period had sustained high wildfire risk in California. We note that the framework is applicable to longer periods, possibly spanning multiple wildfire seasons. We compare the outcomes of the PSPS de-energization framework without fairness consideration (Section~\ref{sec:psps_formulation}) to our proposed framework with fairness consideration (Section~\ref{sec:fairness}). We next present the network and parameters for this test case.


\subsection{Test network}
We demonstrate the PSPS rolling optimization formulation with fairness considerations using a synthetic transmission network: the 73-bus RTS-GMLC test case, Active Power Increase (API) version. This network is geolocated in southern California and parts of Arizona and Nevada. 
The network properties and topology are adopted from~\cite{pglib-opf} based on data originating from~\cite{barrows2019ieee}.
This test network has 73 buses, 99 generators, and 120 transmission lines.
The lower limits of all generators are set to zero to guarantee solution feasibility, i.e., $\underline{g}^i = 0$, $\forall i \in \mathcal{G}$.
We assume a linear routing of lines between their terminal buses.

\subsection{Wildfire risk predictions}
Determining the wildfire ignition risk posed by an energized power line is a challenging task since the risk is dependent on various environmental conditions and the line's physical characteristics \cite{waseem2021}.
Precise calculations of ignition risk values requires detailed data that are not available for our test case and this is not the focus of this work.
Interested readers can see \cite{waseem2021} as well as wildfire mitigation plans published by the California utilities (e.g., \cite{PGE_WildfirePlan,SCE_WildfirePlan}) for more detailed information on calculating wildfire risk.

In this work, we use forecasts of the Wind-enhanced Fire Potential Index (WFPI) produced by the United States Geological Survey as a surrogate for more complex risk-assignment methods~\cite{USGS_WFPI}. WFPI values, which range from 0 to 150, are based on conditions such as wind speed, rain, temperature, etc. Large fires are associated with the highest WFPI values~\cite{USGS_WFPI}. To assign a wildfire ignition risk, $r^\ell$, to each line $\ell \in \mathcal{L}$, we integrate the WFPI index along the line's path. As a result, longer lines have higher risks, which is consistent the risk models used by utility companies~\cite{waseem2021}.




\subsection{Demand prediction and actual values}
For our test case, we consider one day of hourly time periods such that $T=24$. To determine the load demands, we use real hourly load profiles from the California Independent System Operator (CAISO) during the selected wildfire period (June 13 to June 22, 2021)~\cite{CAISO_data}. Assuming that the peak day's highest hourly load during the 2021 wildfire season is equivalent to the nominal demands provided in the test network, we scale the nominal loads in the test case to reflect the real load profiles reported by CAISO.
We incorporate uncertainty in demand forecasts by perturbing the load at each bus for each hour by between $\pm 2$\%, randomly sampled from a uniform distribution. 
These errors are similar in magnitude to those in the load forecasting literature~\cite{hong2016}.

\subsection{Framework parameter values}
The rolling optimization framework includes a number of parameters that we must specify for our test case. We select $\alpha$, the tradeoff parameter used to indicate the system operator's prioritization for reducing the wildfire risk versus reducing the total load shed, based on the forecasted total wildfire risk. We scale $\alpha$ between 0.3 and 0.6, values which were chosen such that a moderate amount of load shedding results for this system. We adjust the $\alpha$ values used for each day according to the total daily wildfire risk prior to any line de-energizations. 
If the total wildfire risk is greater than or equal to that of the highest overall risk seen by the network during the 2019 and 2020 wildfire seasons, then we set $\alpha = 0.3$. If the total risk is less than or equal to the lowest risk seen by the network during these prior years, then $\alpha = 0.6$. Otherwise, we scale $\alpha$ between these two extremes based on the total wildfire risk.

Recall that we also impose constraint~\eqref{const: limit risk} to prevent the solution to \eqref{PSPS-Fair} from significantly increasing the wildfire risk relative to \eqref{PSPS} where we do not consider fairness. The permitted increase in wildfire risk in this constraint is limited to 5\%, i.e., $\zeta=0.05$. We note that there is a substantial uncertainty inherent to assigning energization risk values, so allowing a 5\% increase in overall wildfire risk may be well within estimates of risk uncertainty.

Recall also that we use a forgetting factor $\eta$ to discount the amount of load shed during earlier days of the wildfire period. We select $\eta = 0.9$.
Finally, we evaluate the impact of changing the fairness versus load shedding prioritization by sweeping $\beta$ values from 0.05 to 0.95 in increments of 0.05.



\section{Numerical Results}
\label{sec:results}

As described in Section~\ref{sec:test_case}, we demonstrate our proposed framework by simulating PSPS events over a ten-day period of sustained high wildfire risk in June 2021. We solved the MILPs outlined in Sections \ref{sec:psps_formulation} and \ref{sec:fairness} using Gurobi~9.1.1 to a 1\% MIP gap. 
Optimization formulations were implemented using Julia 1.5.3 with JuMP~0.22.3.
We used the data input functionality of PowerModels.jl 0.19.2~\cite{coffrin2018}.

\begin{figure}
  \centering
  \includegraphics[trim={0 0 2cm 0},clip, width=0.96\linewidth]{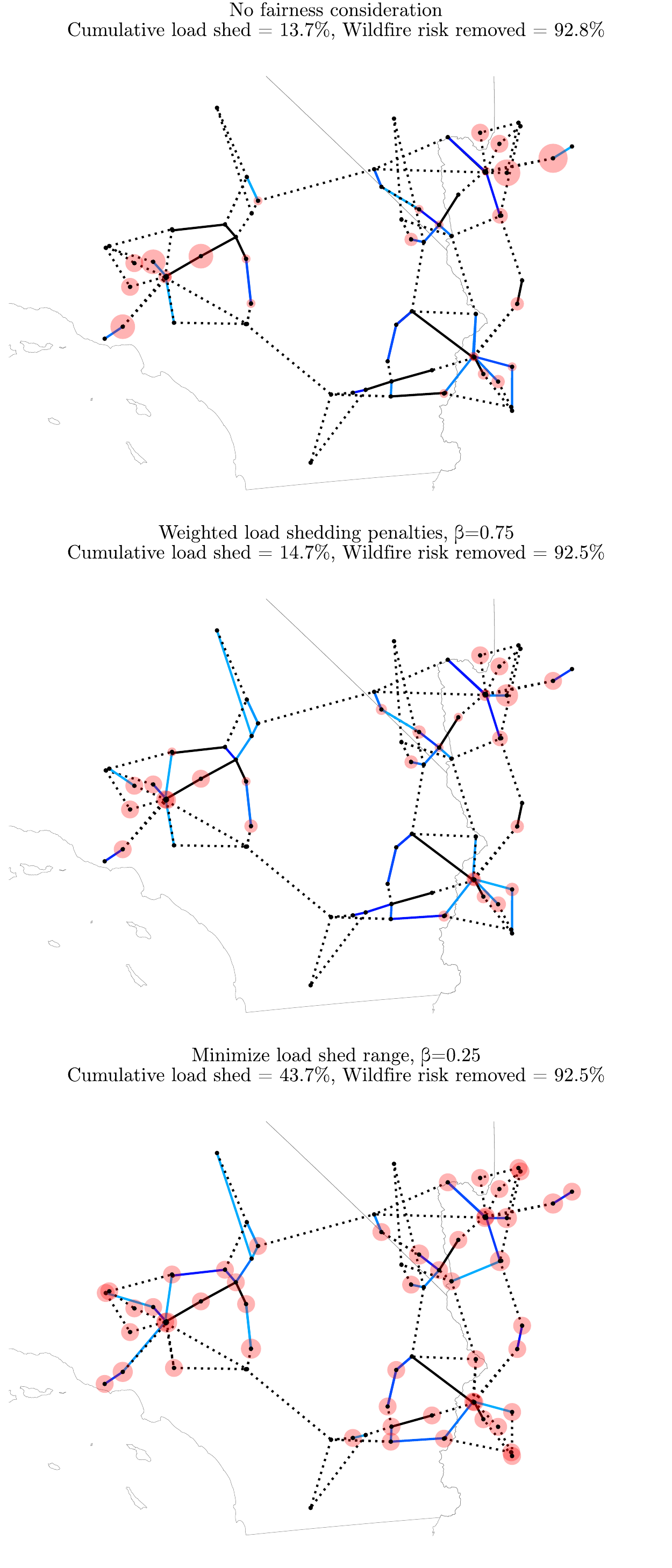}
  \caption{Visualizations of selected optimized PSPS results for a ten-day period of high wildfire risk in June 2021. Each plot shows the RTS-GMLC network. Dots represent buses with the size of the corresponding red circles denoting the amount of load shedding at that bus. Dotted lines are de-energized for every day of this period and solid lines are energized for every day of this period. Blue lines are de-energized during some days, with lighter blue colors denoting lines that are de-energized more frequently.}
  \label{figure:network}
\end{figure}

We illustrate the results with three figures. We first describe these figures and then discuss their implications. Figure~\ref{figure:network} shows three representative cases of the cumulative load shed per bus over the simulation period: (top)~no fairness consideration (Section~\ref{sec:psps_formulation}), (middle) using the weighted load shedding penalties method (Section~\ref{weight}) with $\beta=0.75$, and (bottom) using the minimize load shed range method (Section~\ref{range}) with $\beta=0.25$.
Red circles mark the amount of load shedding at each bus, with larger circles indicating that more load is shed.
Dotted lines are de-energized every day of the simulation period, while black lines remain energized every day.
Blue lines are sometimes de-energized, with lighter blue colors indicating more frequent de-energization.

\begin{figure}[t]
  \centering
  \includegraphics[width=0.95\linewidth]{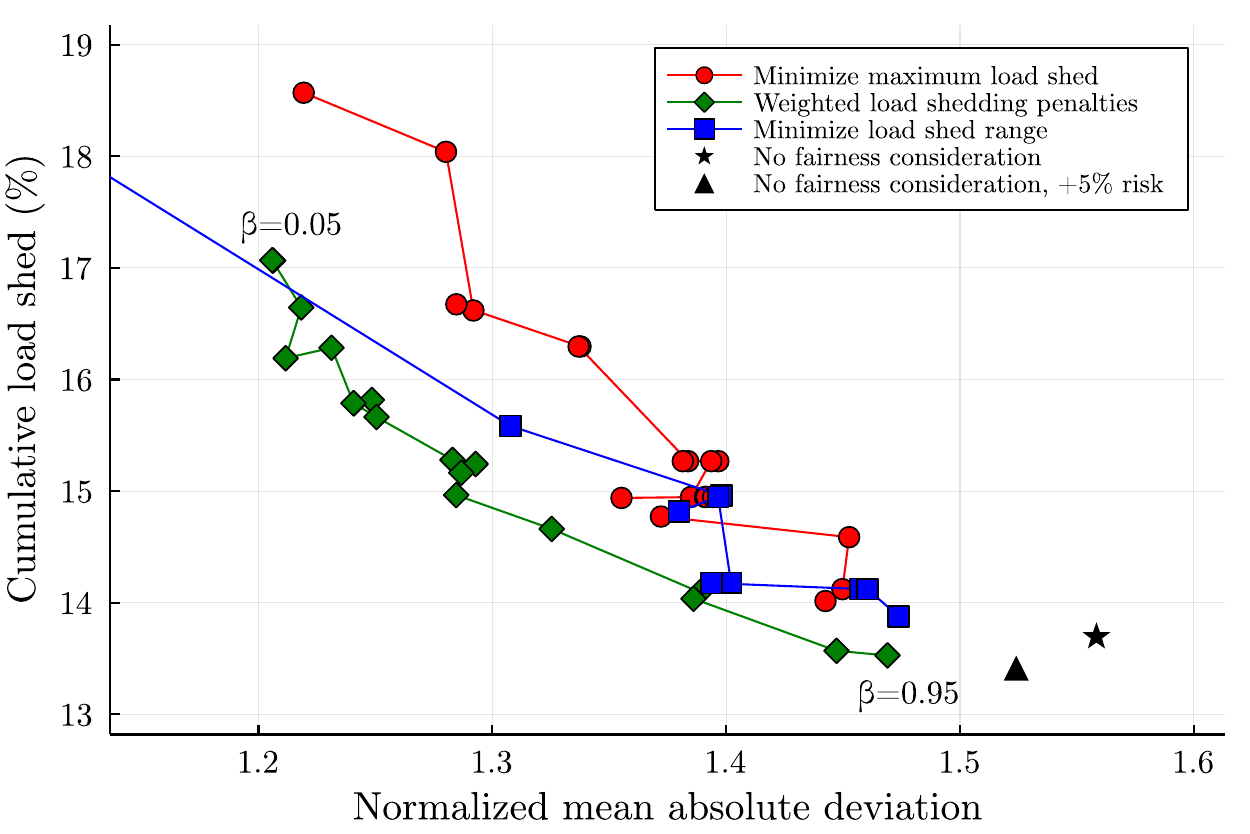}\\
  \vspace{5pt}\includegraphics[width=0.95\linewidth]{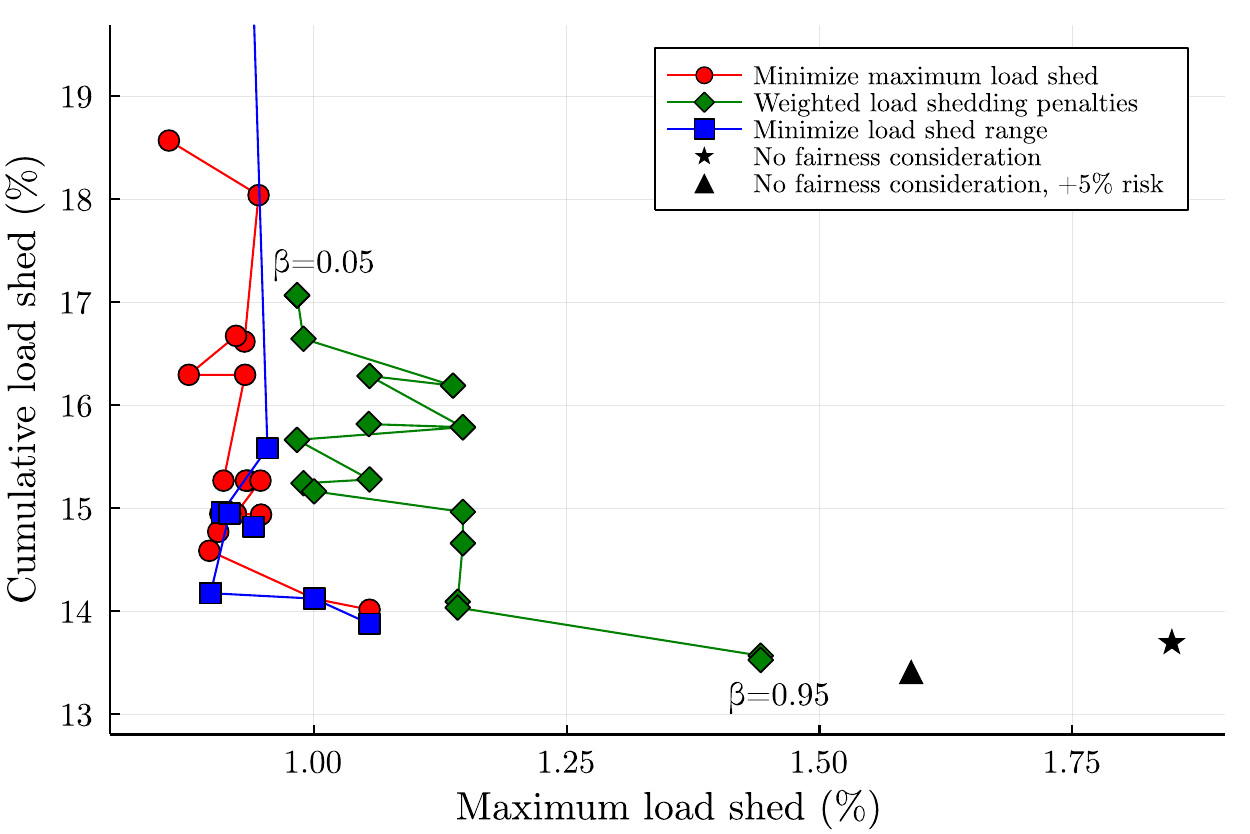}
  \caption{Tradeoff between the cumulative load shed as a percentage of the total demand over the ten-day simulation period and (top plot) the mean absolute deviation of the cumulative load shed per bus normalized by the average cumulative load shed over all buses, or (bottom plot) the maximum load shed over all buses normalized by the total demand in the network over the ten-day simulation. Each line shows the performance of a different fairness method with $\beta$ values ranging from 0.05 to 0.95 in increments of 0.05. The star marks the outcome if fairness if not considered. The triangle marks the outcome if we permit wildfire risk to be up to 5\% greater than the no fairness version. Note that some $\beta$ values for the blue line, which marks the minimize load shed range method, lead to outliers with  large values of total load shedding ($>40\%$) that are not shown in the figure.}
  \label{figure:curves}
  \vspace*{-1.5em}
\end{figure}

Figure~\ref{figure:curves} demonstrates the tradeoff between the cumulative network-wide load shed and fairness while varying the prioritization parameter $\beta$. 
The top plot measures fairness via the mean absolute deviation of the cumulative load shed at each buses normalized by the mean cumulative load shed over the buses.
The bottom plot measures fairness via the maximum cumulative load shed over the buses as a percentage of the total demand in the network. 
The star marks the performance if fairness is not considered (i.e., Section \ref{sec:psps_formulation}). The triangle marks the minimum possible cumulative load shed that can be achieved if the wildfire risk is fixed to be no greater than 5\% of the risk that results from the no fairness method. Thus, the triangle lower bounds the cumulative load shed that could potentially be achieved by any of the fairness methods.

\begin{figure}[t]
  \centering
  \includegraphics[width=0.95\linewidth]{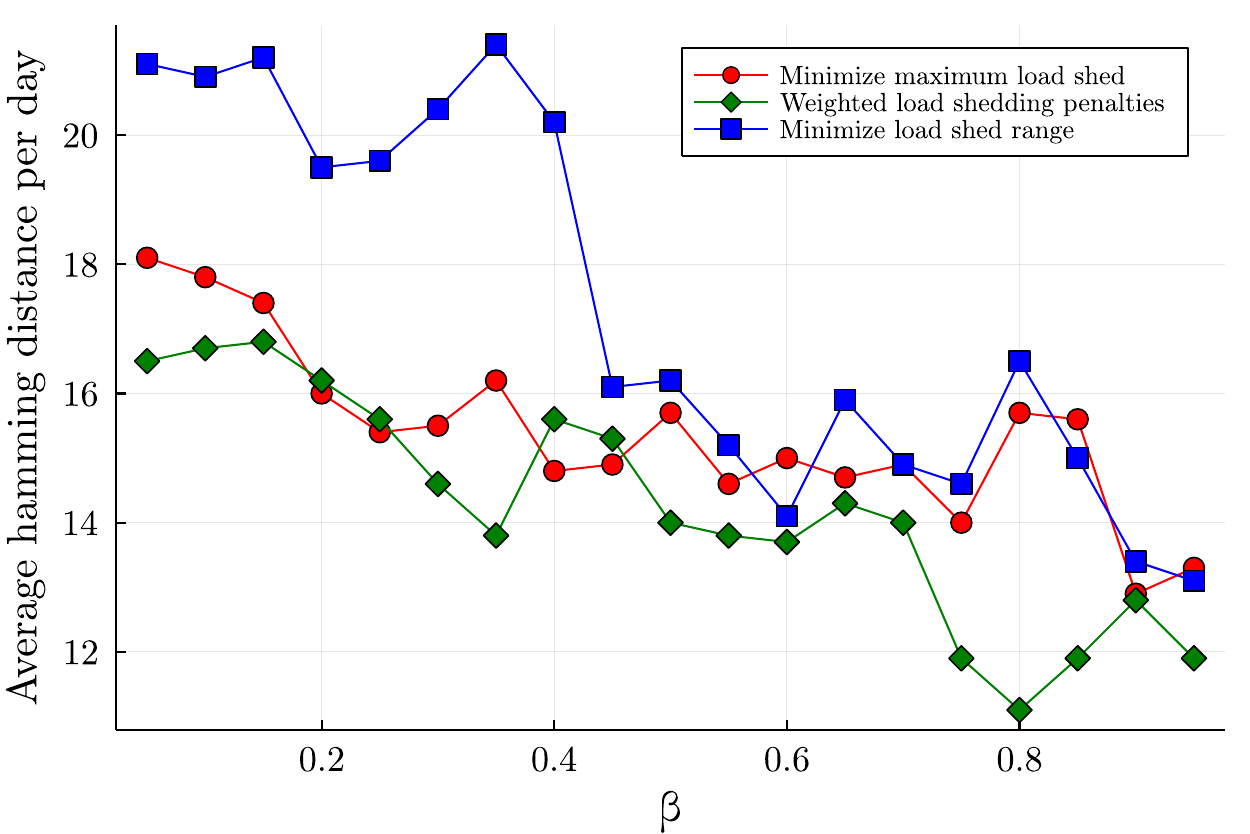}
  \caption{Number of lines with different de-energization decisions from the solutions to~\eqref{PSPS-Fair} for each of the fairness methods, relative to the solution to~\eqref{PSPS} that does not consider fairness (i.e., the Hamming distance between the variables $z$ for these solutions). The values for each day are averaged over the ten-day wildfire period and shown in terms of $\beta$.}
  \label{figure:switch}
  \vspace*{-1.5em}
\end{figure}

Figure~\ref{figure:switch} shows the number of lines with different de-energization decisions in the solutions to~\eqref{PSPS-Fair} versus the solution to~\eqref{PSPS} (i.e., the \emph{Hamming distance} between the variables $z$ for these solutions), averaged over the days in the wildfire period for varying values of $\beta$. 

Comparing the top and middle plots in Figure~\ref{figure:network} shows that considering fairness leads to a more even distribution of load shed among the buses in the network.
This more even distribution comes at a limited cost (an additional 1\% cumulative load shed and a 0.3\% increase in wildfire risk relative to the problem that does not consider fairness). Figure~\ref{figure:curves} reinforces this observation with the results for large values of $\beta$ (prioritizing reductions in total load shedding) showing that substantial fairness improvements can be achieved with limited tradeoffs in the total load shedding. (See the lower portions of the curves in Figure~\ref{figure:curves} with $\beta = 0.95$ relative to the triangle denoting no fairness considerations.)

However, more strongly prioritizing fairness can lead to erratic behavior as shown by larger values of $\beta$ in Figure~\ref{figure:curves}. Some of this behavior may be attributed to the inherent discontinuous nature of discrete de-energization choices and the maximization functions in the first and third fairness methods, and the load uncertainty also contributes. However, we hypothesize that much of this behavior is due to the sequential de-energization decisions made daily in our framework. Extending~\eqref{PSPS-Fair} to consider wildfire risk forecasts over a multi-day horizon may both improve the quality of the solution and smooth this erratic behavior. Such extensions are the focus of our ongoing work.

Since the RTS-GMLC system has numerous generators and a very robust network, many lines can be de-energized while still supplying most loads. While fewer lines would likely be de-energized for actual systems in practical settings, this test case still gives useful insights. Our ongoing work includes scaling the framework to larger, more realistic cases.

We also observe from the top and middle plots in Figure~\ref{figure:network} that line switching decisions are different when incorporating fairness, meaning that there is a benefit to jointly considering de-energization and fairness as opposed to first selecting lines to de-energize and then re-dispatching generation to maximize fairness. 
The large number of different line de-energization decisions shown in Figure~\ref{figure:switch} emphasizes this observation and further illustrates that the fairness methods can produce significantly different results, especially when prioritizing fairness via small values of $\beta$.

Finally, we note that the bottom plot in Figure~\ref{figure:network} shows an example of a perverse outcome if fairness considerations are not carefully designed.
Here, we see that the red load shedding circles are uniformly sized, meaning that load shed is more evenly distributed among the buses; however, the cumulative load shed is 30\% greater than is obtained without considering fairness. We hypothesize that the solver increases the load shedding at certain buses beyond what is necessary to reduce the maximum load shed (i.e., to reduce the range of load shedding in the network, the solver chooses to increase load shedding at all buses).
This is clearly an undesirable outcome since, all else being equal, we would always prefer to supply more load if it were possible to do so without increasing the amount of load shed for another customer.
Although this load shed range method provides more reasonable outcomes for higher values of $\beta$, this scenario demonstrates that undesirable outcomes may occur if a fairness method is not appropriately designed.

\section{Conclusions}
\label{sec:conclusion}
Increasingly frequent and severe wildfire conditions driven by climate change motivate the development of new computational tools for efficiently and fairly executing PSPS events to mitigate acute wildfire ignition risks. The rolling optimization framework proposed in this paper determines line de-energization choices that optimize system-wide performance with respect to wildfire ignition risks and total load shedding as well as fairness considerations for the load shedding at each bus. We analyzed three different methods for promoting fairness in the load shedding: minimizing the maximum load shed at any bus, minimizing weighted load shedding penalties, and minimizing the range of load shedding across buses. Based on numerical demonstrations of this framework using the RTS-GMLC test case with actual wildfire risk and load profile data, we emphasize three key observations. First, with appropriate selection of the tradeoff parameter $\beta$, the framework can achieve significantly more fair outcomes with limited increases in both the wildfire risk and the total load shedding. Second, the method chosen to promote fairness matters since the framework gives significantly different outputs with respect to both the load shedding at each bus and the line de-energization decisions. Third, certain methods for promoting fairness may produce perverse incentives, as minimizing the range of load shedding across buses led the solver to unnecessarily increase the load shed at some buses.

These observations motivate our ongoing work in formulating and analyzing alternative methods for promoting fairness in optimization problems related to wildfire risk mitigation. We particularly intend to study the computational characteristics of different methods for considering fairness. Using stochastic optimization techniques, we also aim to generalize the proposed rolling optimization framework to consider multi-day horizons in order to obtain higher-quality solutions while addressing forecast errors for the wildfire risks, renewable generation, and load demands. We additionally plan to study the impacts of more realistic formulations that consider contingencies, anti-islanding, and re-energization~\cite{rhodes2022co}. Finally, we intend to extend this paper's work on fairness to consider equity by investigating the impacts of PSPS events on different populations.






\section*{Acknowledgement}
We gratefully acknowledge insightful discussions on this topic with N.-J.~Simon. We also note that this research was supported by resources provided by the Partnership for an Advanced Computing Environment (PACE) at the Georgia Institute of Technology. We appreciate assistance from R.~Piansky on setting up the numerical tests on PACE.

\bibliographystyle{IEEEtran}
\bibliography{IEEEabrv,references_final_submission.bib}

\end{document}